\documentclass[aps,prl,twocolumn,floats,showpacs]{revtex4}
\usepackage{graphicx}
\begin{document}
\title{Shot noise in the interacting resonance level model}
\author{ A. Golub}
\affiliation{ Department of Physics, Ben-Gurion University of the
Negev, Beer-Sheva, Israel \\   }
\begin{abstract}
We calculate the shot noise power and the Fano factor of a
spinless  resonant level model. The Coulomb interaction which in
this model acts between the lead electron and the impurity is
considered in the first order approximation. The logarithmic
divergencies which appeared in the expressions for shot noise and
the transport current are removed by  renormalization group
analysis. It is important, that unlike a few recent statements,
Keldysh technique gives an adequate description of perturbation
theory results. By passing to the bosonized form of the resonance
model we show that in the strong interaction limit the tunnelling
becomes irrelevant and decreases.
\end{abstract}

\maketitle {\it Introduction}: During the last two decades the
transport properties of quantum impurity  have attracted great
interest as a basic problems in nanophysics \cite{gold}. Recently,
spinless interacting resonance level model (IRLM) \cite{gogolin}
has become a subject of a special attention
\cite{andrei,borda,doyon,saleur,nishino}. This interest, in
particular, was related to the extending the Bethe ansatz out of
equilibrium to calculate the transport current in the steady state
of a quantum dot \cite{andrei}. In this work  IRLM, particulary,
served as a simplest test and toy model. The Mehta and Andrei
showed that the transport current in the interacting resonance
level model is changed non monotonically with the strength of
Coulomb coupling. The analysis of this result in reference
\cite{borda} was provided by time-ordered scattering formalism in
the next-leading logarithmic order and, indeed, it was  found a
correction to the current in \cite{andrei}. Close to the work
\cite{andrei} a new method to calculate the transport current at
the first order in Coulomb interaction was suggested by Doyon
\cite{doyon}. This method takes into account the jump of the wave
function at the impurity site (impurity condition) and exploits
the Hershfield's approach \cite{hersh} to nonequilibrium dynamic.
Our formula for the transport current slightly deviates from his
result and we also have obtained the current more directly
 with the help of Keldysh technique. The bosonization of
IRSM hamiltonian made in \cite{saleur} was used to calculate
linear conductance in the region for Coulomb coupling strength
where the tunnelling remains relevant. The bosonized form of the
hamiltonian permits a clear identification of this region and
shows that for sufficiently large  Coulomb coupling the
hybridization energy flows  to zero.

 The principal
purpose of this work is the calculation of the current-current
correlation function, in particular, the zero frequency shot noise
power-the value This value is very important experimentally
measured characteristic and technically it is much more difficult
to obtain. Shot noise has not be considered in the cited articles
and this value, probably, causes a special difficulties for Bethe
ansatz approach in nonequilibrium. Here we obtain the shot noise
power for a general position of bear impurity level $\epsilon_d $
relative to the Fermi energy which is taken at zero.

{\it The hamiltonian}: The standard form of the hamiltonian of
IRLM has a form
\begin{eqnarray}
H&=&H_0+t(\sum_i \psi_i(0)d^{\dagger}+h.c)+\epsilon_d d^{\dagger}d
\\ \nonumber
 && +U\sum_i
\psi_i(0)^{\dagger}\psi_i(0)d^{\dagger}d \label{h}
\end{eqnarray}
The first  term corresponds to non-interacting electrons in the
two leads
\[H_0=\sum_{ik}\varepsilon_{i}(k)\psi_{ik}^{\dagger}\psi_{ik}\]
 where $\psi_{ik}$, $ \varepsilon_{i}(k)$
are the electron field operator  and the electron energy of a lead
$i$, respectively. Index $i=L,R$ indicates left (right) lead. The
second term describes the tunnelling processes. The last one
presents an important interacting term with Coulomb capacitive
coupling $U$. For non-equilibrium transport properties we use
Keldysh technique and go to the action generalized for Keldysh
space. We develop a perturbation theory in $U$, therefore,
separate the action $A$ into two parts $A=A_0 +A_U$ where
\begin{equation}\label{o}
  A_U =-U\int dt[\sum_i
\psi_i(0)^{\dagger}\psi_i(0)d^{\dagger }d]^p \sigma_z^{pp}
\end{equation}
here $\sigma_z $ stands for Pauli matrix,  the summation over
repeated Keldysh indices $p=1,2$ is performed. Also $[..]^p$ means
that to every operator in the brackets we have to assign Keldysh
index $p$. The remaining part of the action is easily diagonalized
by a shift of electron operators in the leads
\begin{eqnarray}
\psi_{ik}^{\dagger p}=\tilde{\psi}_{ik}^{\dagger p}+t
(d^{\dagger}\sigma_z g_{ik})^p &, &\psi_{ik}^{
p}=\tilde{\psi}_{ik}^{ p}+t ( g_{ik}\sigma_z d)^p \label{shift}
\end{eqnarray}
Here $g_{ik}$ are the matrix Keldysh Green functions (GF) of the
 electrons with energy $\epsilon(k)$ in the lead $i$. Actually we need GF at the impurity site
$x=0$: $g_i(0)=2\pi N(0)g_i $. Here N(0)is the density of states
in the wire and $g_i=\int d \epsilon_k g_{ik}$. The Keldysh
rotation $W=\frac{1}{\surd 2}\left(\begin{array}{cc}
  1 & -1 \\
  1 & 1
\end{array}\right)$
leads to a especially simple form of these functions:
$g_n'=W\sigma_z g_n W^{-1}$
\begin{equation}\label{prime}
    g_n '= \left(\begin{array}{cc}
 g_n ^R & g_n ^K \\
         0 & g_n ^A
\end{array} \right)
\end{equation}
where $g_n^{R,A}=\mp i/2;g_{L,R}^K(E)=-i\tanh[(E\pm eV/2)/2T]$,
  $T$ is the temperature and $V$ -the applied voltage. Below we
 consider the case of $T=0$. With the help of (\ref{shift}) the
 action $A_0$ then becomes

\begin{eqnarray}
A_0&=&\int\int dt
dt'[\sum_{ik}\tilde{\psi}_{ik}^{\dagger}g_{ik}^{-1}\tilde{\psi}_{ik}+
d^{\dagger}G^{-1} d] \label{A0}
\end{eqnarray}
here  modified Keldysh GF of impurity level has appeared. After
Keldysh rotation (\ref{prime}) this GF transforms to $G'$
\begin{equation}\label{G}
  G^{'-1}=G_0 ^{'-1}-\Gamma(g'_L+g'_R)
\end{equation}
where $\Gamma=2\pi N(0)t^2 $ stands for hybridization energy and
$G_0$ is the GF of a bear noninteracting level.

{\it Shot noise}: The current noise power is given by
current-current correlation function
$S_{tt_1}=1/2(<I(t)I(t_{1})>+<I(t_{1})I(t)>)-<I>^2$ which in the
Keldysh time ordering form can be rewritten as
\begin{equation}\label{SG}
    S_{tt'}=1/2(<\hat{T}I^2 (t)I^1(t')>+<\hat{T}I^1 (t)I^2 (t')>)
\end{equation}
Here the superscripts $1,2$  are related to the branches of the
time contour. Current operator is obtained by commutation of the
density operator of the left or right lead with the hamiltonian
(\ref{h}). In a symmetric presentation the current operator
acquires a form
\begin{equation}\label{j}
  I=\frac{ite}{2}<\hat{\psi}(0)^{\dagger}d-d^{\dagger}\hat{\psi}(0)>;
\hat{\psi}=\psi_L-\psi_R
\end{equation}
Next we insert (\ref{j}) into formula for noise (\ref{SG}) and
obtain now  a working expression to which the Keldysh technique is
applied. In this formula the average is performing with the total
action $A_0+A_U$. Considering the shift (\ref{shift}), we, at
first, calculate the averaging of equation (\ref{SG}) with the
action $A_0$.  As a result we get
 transport current $I_0=\frac{e\Gamma}{2\pi}j_-$ and shot
 noise $S_0$ for
noninteracting resonance level
\begin{eqnarray}
S_0&=&\frac{e^2\Gamma}{4\pi}(j_{-} -j_1)\\
j_1&=&\Gamma[\epsilon_{+}(\epsilon_+^2 +
\Gamma^2)^{-1}-\epsilon_{-}(\epsilon_-^2 + \Gamma^2)^{-1}]
\end{eqnarray}
where here and below
$j_{\pm}=\arctan\frac{\epsilon_{+}}{\Gamma}\pm\arctan\frac{\epsilon_{-}}{\Gamma}$
and $\epsilon_{\pm}=\epsilon_d \pm eV/2$. In the linear regime
(small voltages: $V/(\epsilon^2+\Gamma^2)<<1$) we immediately
recognize standard expressions $I_0=(e^2/h)\bar{T}$,
$S_0=(e^3/h)\bar{T}(1-\bar{T})$, where
$\bar{T}=\Gamma^2/(\Gamma^2+\epsilon_d ^2)$ is the transmission
coefficient. In the last formulas we have returned to the standard
units (here $2\pi\rightarrow h$).

 To consider the  Coulomb
interacting contribution we expand the average in the noise
(\ref{SG}) and current formulaes to the first order in $A_U$. It
is interesting to notice that in the Keldysh approach the only
divergencies for the shot noise and the transport current are of
logarithmic type which can be accounted by using the
renormalization group. Again performing the shift of leads
operators (\ref{shift}) we result in a sum of time ordering
products for $\psi_i^j$ and $d^j$ operators which have a simple
diagrammatic presentation. After rather long, though, direct
calculations we arrive at the shot noise power $S=S_0 +S_U$ where
\begin{eqnarray}
S_U &=& \frac{u\Gamma}{4\pi^2} \{ [j_{-}-3j_1-
\Gamma^2\frac{\partial (j_1 /\Gamma)}{\partial \Gamma}]L+j_2
j_{+}(1-j_{3}) \} \nonumber\\
j_2&=& \Gamma^2[(\epsilon_+^2 + \Gamma^2)^{-1}-(\epsilon_-^2 +
\Gamma^2)^{-1}] \label{SU}\\
 j_3&=&\Gamma^2[(\epsilon_+^2 +
\Gamma^2)^{-1}+(\epsilon_-^2 + \Gamma^2)^{-1}]\nonumber\\
L&=&\ln[\frac{D}{[(\epsilon_+^2 + \Gamma^2)(\epsilon_-^2 +
\Gamma^2)]^{1/4}}]\nonumber
\end{eqnarray}
Similarly, the capacitor coupling leads to a contribution to the
transport current
\begin{eqnarray}
I_U&=&\frac{u\Gamma}{2\pi^2}\{(j_- -j_1)L+
\frac{\pi}{8}\ln\frac{\epsilon_-^2 + \Gamma^2}{\epsilon_+^2 +
\Gamma^2}+\frac{1}{2}j_2 j_+\}\label {ju}
\end{eqnarray}
In equations (\ref{SU}) and (\ref{ju}) $D$ is the large cut-off of
the order of band width, and we also introduced the notation
$u=2\pi U N(0)$. It is straightforward to prove by direct check
that the shot noise power, the same as the current \cite{doyon},
satisfies the same Callan-Simanzik equation with  $\beta$-function
$\beta=-u\Gamma/\pi$
\begin{equation}\label{CS}
    D\frac{\partial S}{\partial D}+\beta\frac{\partial S}{\partial
 \Gamma}=0
\end{equation}
Renormalization group  equation for hybridization energy follows
immediately $\mu\frac{\partial \Gamma(\mu)}{\partial
\mu}=-\frac{u}{\pi}\Gamma $ where $ \mu $ is a physical cut-off.
This equation  can be solved in the scaling limit by introducing a
parameter $T_K$. This parameter, like Kondo temperature, can be
defined at the scale $\mu=D$ and the bear value of $\Gamma$: $T_K
^{1+u/\pi}=\Gamma D^{u/\pi}$. The solution gives hybridization
coupling as function of physical cut-off $\mu$
\begin{equation}\label{gamma}
    \Gamma(\mu)=T_K
^{1+u/\pi}(\mu)^{-u/\pi}
\end{equation}
There is one important remark. Let us use definition of $T_K$
 in terms of high energy cut-off $D$ and
bear tunnelling width (as given above).  For repulsing interaction
($U>0$) we  get an increase of $\Gamma(\mu)$ and, correspondingly,
the increase of the average current. This happened because $D$ is
the largest scale $D>\mu$. However, in the universal region we can
properly fix the scale $T_K$  and obtain the decrease of the
current as $U$ grows.
\begin{figure}
\begin{center}
\includegraphics [width=0.45 \textwidth ]{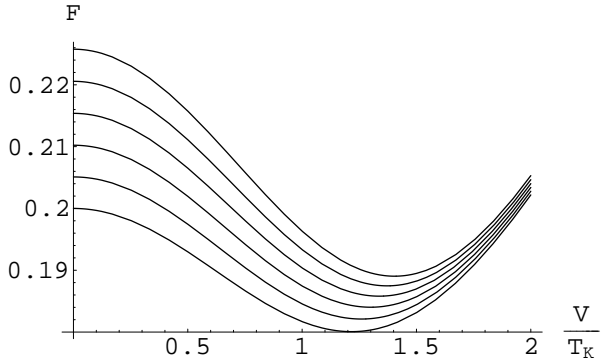}
\caption { Fano factor as function of applied voltage for
different values of $u$. The numerical values of chosen
parameters: $\mu/T_K= 3, \epsilon_d/T_k=-0.5.$ The Coulomb
coupling $u/\pi$ is changed
 from 0 (lowest curve) up to $u/\pi$=0.1}
\end{center}
\end{figure}
Even at $U$=0 the shot noise is a nonlinear function of hopping
probability. In this case the Fano factor $F=S/I$ reaches the
maximum value at $\bar{T}=1/2$, and $F=0$ at resonance position of
the level ($\epsilon_d =0$). In Fig1. we plot the Fano factor as a
function of voltage for different values of Coulomb coupling $U$.
The lover curve correspond to $U=0$. As to the physical cut-off
parameter $\mu$, it is smaller then $D$, actually $\mu<<D$. In the
scaling renormalised regime moderate change of $\mu$ has, indeed,
little impact on the noise value.

{\it Strong Coulomb interaction}: The repulsive case for large $U$
was investigated by renormalization group approach \cite{borda}
with a conclusion that
 that the orthogonality catastrophe reduces the
hybridization energy considerably. Here we use \cite{saleur} the
bosonization of the hamiltonian (\ref{h}) as a method, which is
especially convenient for application to the present problem.
After unfolding and linearizing near the Fermi surface, one ends
up with the hamiltonian $H_0$  for two chiral right movings
electrons $R_n$ (n=1,2). Next, applying the unitary transformation
$Y=\exp[ieVt/2 N_V]$ ($N_V =\int dx R^{\dagger}_j \tau_z ^{jj} R_j
$) \cite{glazman} we move all dependencies on the applied voltage
to the tunnelling part of Hamiltonian. Recalling the form of the
current operator it is naturally to include the current as a
source part $ \alpha(t) \hat{I} $ to an effective tunnelling
hamiltonian
\begin{eqnarray}\label{ht}
  \tilde{H_T}&=&t[(1+i \alpha)R_1 (0)e^{-ieVt/2}+\nonumber\\
  &&(1-i \alpha)R_2(0)e^{ieVt/2}]d^{\dagger}+h.c
\end{eqnarray}
We write the hamiltonian  in terms of even and odd combinations
$R_{\pm}=(R_1 \pm R_2)/\sqrt {2}$. Let us bosonize
$R_{\pm}=\frac{\eta_{\pm}}{2\pi}e^{i \sqrt {4\pi} \varphi_{\pm}}$.
This yields
\begin{eqnarray}\label{hb}
  H&=&\sum_{n=\pm}H_0 (\varphi_{n})+\frac{u}{\sqrt{\pi}}
  (\partial_x \varphi_{-}(0)+\partial_x
  \varphi_{+}(0))d^{\dagger}d+\nonumber\\
  &&\frac{t}{\sqrt{\pi}}[\eta_{+}c(t)e^{i \sqrt{4\pi}
  \varphi_{+} (0)}+
 ia(t) \eta_{-} e^{i \sqrt {4\pi}
 \varphi_{-}}]d^{\dagger}+h.c\nonumber
\end{eqnarray}
Here $H_0 (\varphi_{n})=\frac{1}{2}\int dx( \partial_x
\varphi_{n})^2 $. The time dependent functions $c(t), a(t)$ have a
form $ c(t)=\cos(eVt/2)+\alpha(t)sin(eVt/2)$ and $
a(t)=\alpha(t)\cos(eVt/2)-sin(eVt/2)$. If we pass to the action on
Keldysh time contour then the source function $\alpha $ is
transformed into a diagonal matrix with the entries $\alpha_1$ and
$\alpha_2$. Next, to cancel the direct Coulomb interaction we
perform the unitary transformation
$\hat{K}=e^{i\frac{u}{\sqrt{\pi}}d^{\dagger}d(\varphi_{-}+\varphi_{+})(0)}$.
At this stage, we use the new canonical boson variables
\cite{saleur}
$\vartheta_{\pm}=[(\sqrt{4\pi}-\frac{u}{\sqrt{\pi}})\varphi_{\pm}-
\frac{u}{\sqrt{\pi}}\varphi_{\mp}]/\beta$ with
$\beta=\frac{2}{\pi}(u-\pi)^2 +2\pi $ and transform the
hamiltonian into the desired form convenient for constructing the
Keldysh effective action.
\begin{eqnarray}
  H&=&\sum_{n=\pm}H_0 (\vartheta_{n})+\frac{t}{\sqrt{\pi}}[c(t)\eta_{+}e^{i\beta
  \vartheta_{+}(0)}+\nonumber\\
 &&ia(t)\eta_{-}e^{i \beta (1-4\pi/\beta^2)\vartheta_{+}+
 i(4\pi/\beta)(1-u/\pi)\vartheta_{-}}]d^{\dagger}+h.c\nonumber
\end{eqnarray}
The effective action with this hamiltonian can be used also in the
 limit of small $U$. For this we expand the dimension parameter $\beta$ to first
 order in $U$ and return to fermion representation. Here we
 notice  that the whole perturbation now is related to the
 tunnelling hamiltonian. The renormalized hopping rate becomes
 irrelevant perturbation if $\beta^2 > 8\pi$ or $|U-\pi|>
 \sqrt{3}\pi$.  In this case the nonlinear dependence of shot
 noise power on transmission rate is not important and, thus, Fano factor
$F\rightarrow 1$.

{\it Details of calculations}: Our principal equation (\ref{SU})
as well as the equation for the transport current follow as a
result of summations of a series of Keldysh diagrams representing
contributions  of the first order in $U$. We get general
expression for $S_U$  by inserting the action $A_U$ and the
current operators given by eq.(\ref{j}) under the average in
(\ref{SG}).
\begin{eqnarray}
S_U&=&\frac{iUt^2e^2}{4}<\hat{T}\int dt_{1}
(\psi_i(0)^{\dagger}\psi_i(0)d^{\dagger }d)_{t_1 }^p
\sigma_z^{pp}(\hat{\psi}(0)^{\dagger}d\nonumber\\
&&-d^{\dagger}\hat{\psi}(0))_{t}^2 (\hat{\psi}(0)^{\dagger}d-
d^{\dagger}\hat{\psi}(0))_{t'}^1> \label{SUir}
\end{eqnarray}
where the summations over repeated indices is implied.  Recalling
the shift (\ref{shift}) we can express (\ref{SUir}) only in terms
of level $d$ and tilde leads operators. All operators now are
written in the interaction representation relative the quadratic
action $A_0$. After performing transformation (\ref{shift}) shot
noise power $S_U$ is separated into a three blocks having,
correspondingly, four, two and none of tilde electron operators.
As a representative example we consider here the simplest black
with two pairs of tilde operators (like $\tilde{\psi}_l
(0)\tilde{\psi}^{\dagger}_i(0)$). Due to Wick theorem with
quadratic action $A_0$ the contractions of electron's and impurity
$d$ operators are separated and we end up with four diagrams
presented on Fig.2.
\begin{figure}
\begin{center}
\includegraphics [width=0.4 \textwidth ]{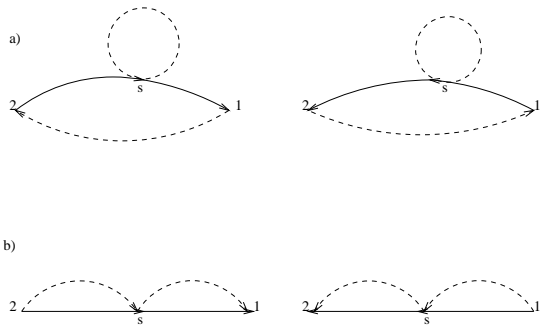}
\caption { Keldysh diagrams which contribute to $S_U ^ {'}$. Solid
lines represent the leads electron GF $g_i^{\alpha,\beta}$, the
dash ones stand for dot GF $G(\epsilon)$. The numbers $1,2,s$ are
the Keldysh  time indices. The topologically similar diagrams
give, nevertheless, different contributions due to rearrangement
of Keldysh indices}
\end{center}
\end{figure}
Explicitly, only two diagrams $(b)$ give nonzero contribution. We
can write this contribution as $S_U ^{'}=\frac{iu\Gamma}{4}Q_U$,
where
\begin{eqnarray}
Q_U&=&\sum_{i\alpha}\int\frac{d\epsilon}{2\pi}G^{2
\alpha}(\epsilon)g_i ^{\alpha 2}(\epsilon)\sigma_z ^{\alpha
\alpha}\int\frac{d\epsilon'}{2\pi}G^{\alpha 1}(\epsilon')g_i ^{1
\alpha}(\epsilon') \nonumber \\
&& +(1 \Leftrightarrow 2) \label{QU}
\end{eqnarray}
 Green's functions $G$ and $g_i$ in Eq.(\ref{QU}) are related to  Keldysh GFs $g',G'$ (see
(\ref{prime}), (\ref{G})). In particular,
$G^{R,A}(\epsilon)=(\epsilon-\epsilon_0\pm i\Gamma)^{-1}$ and
$G^K=G^R (g_L ^K+g_R ^K)G^A$. In the limit of zero temperature he
energy integrations can be easily performed with the result
\begin{equation}\label{Q}
  Q_U=-\frac{i}{\pi^2}Lj_{-}
\end{equation}
The other blocks of diagrams include more functions $g_i$ and $G$,
however, the energy integrations are simple and can be easily
done.

 {\it Conclusion}:
 We used standard Keldysh technique to calculate for the first
 time zero frequency shot noise power in the limit of zero
 temperature in interacting resonance level model.
  We have shown that in spite  of existing scepticism against
 the application of Keldysh theory
to nonequilibrium stationary interacting problems, at least, for
the weak Coulomb coupling in IRLM this theory is justified. In the
strong interacting limit we construct effective action based on
bosonization of the Hamiltonian. We was shown that the
orthogonality catastrophe causes the Fano factor to approach one.
It is important that the Coulomb coupling is not renormalized,
($U$ does not flow), therefore, renormalized perturbation theory
 is correct for all value $\epsilon_d,V,T_K $ $\ll$$D$.

\begin{acknowledgments}
This work was inspired by the  lectures given by N. Andrei on
Bethe ansatz in nonequilibrium.
 I would like to thank B. Horovitz and Y. Meir, for discussion.

\end{acknowledgments}

\end{document}